# Identification of phases in scale-free networks

S. Jenkins and S. R. Kirk

*Dept. of Informatics & Mathematics, HTU, Trollhättan, 461 29 SWEDEN*

PACS. 05.10. –a – Computational methods in statistical physics.

PACS. 05.65. +b – Self-organising systems.

**Abstract.** There is a pressing need for a description of complex systems that includes considerations of the underlying network of interactions, for a diverse range of biological, technological and other networks. In this work relationships between second-order phase transitions and the power laws associated with scale-free networks are directly quantified. A unique unbiased partitioning of complex networks (exemplified in this work by software architectures) into high- and low-connectivity regions can be made. Other applications to finance and aerogels are outlined.

The huge pressure on software development teams for ever-faster turnarounds on software specifications with decreasing budgets, and the increasing use of distributed computing both highlight the need for cheaper testing methods. A recent article [1] addresses the topic of understanding the costs of defects present in software, both before and after release. It was found that (a) finding and fixing a *severe* software problem is often 100 times more expensive after delivery than during the requirements and design phase, and (b) about 80% of the wasted effort comes from 20% of the defects. Further to this (c) it was also found that 'about 90% of the downtime comes from at most about 10% of the defects' [1]. Finally (d), [1] states that for operational systems, only about 2% of defects recorded caused the system to go down (i.e. so-called 'category 1 defects', where the system was "Dead in the Water"). Point (a) above demonstrates that it is extremely important to be able to find the damaging defects (*i.e.* those causing the most widespread damage), and Point (b) highlights the importance of efficiency in finding these defects. Point (c) illustrates, most relevantly to this work, that there is something very time consuming about the process of fixing defects in a small fraction (i.e. probably less than 10%) of the software modules. We demonstrate in this work that an explanation for Points (c) and (d) above, (i.e. the large and inefficient amount of downtime), is provided by a quantitative understanding of the structure of the network of the software modules, and how that relates to testing, in particular retesting efforts.

The understanding of the importance of the structure of the relationships between classes (or more generally, functional units) in software leads to the realization that a selective software testing strategy may be more effective than a conventional random testing approach, with the consequence that excessive time and expense either before or after product release could be avoided. In this work we take a pragmatic view of errors in software; there have always been errors present in software and it is likely that there always will be. The difference in our approach is that errors need not all be sought with the same urgency in the timeline of a project. We attempt to understand the effect of the location of errors on testing and subsequent debugging effort. We can represent software as a network of interconnected nodes [2], where the nodes represent classes in an object-oriented system [3], or, more generally, functional units. The set $Q$ (see later) contains the most highly-connected nodes of the network; removing the errors on these nodes first, at each stage of testing, is the most efficient way to proceed. Since the disruptions caused by addressing defects located on the most highly connected nodes will be greatest, it makes sense to minimize the disruption of the structure of the code as the time goes by in the testing process, in order to expedite the necessary 'in-place' retesting. This work cannot make predictions about the threshold for acceptable performance of a software system in terms of bugs - this depends on the particular application and the tolerance of the users. The software engineer already realises that there will be errors randomly distributed throughout the software nodes, but the question of where to start searching can be tackled by understanding the connectivity of the software.

During the 'alpha testing' stage of software development, prioritised removal of errors on highly-connected nodes will make the testing process run much more smoothly than would be the case with random [4-11] error testing. The 'beta-testing' could then proceed whilst the errors in nodes outside the set $Q$ are removed. In 'Extreme Programming' (XP) development strategies [12], the customer can make more informed decisions at the acceptance stage of testing. The customer specifies scenarios to test when their required functionality has been correctly implemented. A software development project can have as many acceptance tests as are needed to ensure the intended functionality is present. In this stage of testing, customers take responsibility for verifying the correctness of the acceptance tests and reviewing test scores to decide which failed tests are of highest priority. Using our approach, prioritisation is performed automatically, based on a quantitative analysis of the network structure of the software.

A complex network can be characterized by a distribution function $P(k)$, which gives the probability that a randomly selected node has exactly $k$ incident edges. This work shows that the network of software class dependencies (as is the case for most large networks, such as the World Wide Web [13], the Internet [14] or metabolic networks [15]) has a degree distribution which differs significantly from the Poisson distribution characteristic of a random graph (with a peak at $P(<k>)$). In contrast, the degree distribution possesses a power-law tail $P(k) \sim k^{-\gamma}$, with $\gamma > 2.0$. Networks having this type of distribution are known as 'scale-free' [16] networks, due to the lack of a characteristic scale. It is known that scale-free networks display a very high degree of tolerance against random failures [17]. The downside of this property is that scale-free networks are very vulnerable to the loss of a few very highly connected nodes. Some artificial networks (such as those found in software architectures), however, have a restriction that natural networks do not; namely that failure of any single node may cause the failure of the entire network, i.e. there is no homeostasis against random failure.

Within the theory of complex networks there have been a number of papers examining the subject of phase transitions; for example, describing how an entire network moves from one phase to another, demonstrating behaviour similar to a physical Bose-Einstein condensate [18]. In [18] nodes correspond to energy levels and the edges represent particles. In this paper we also employ parallels with thermodynamics, but with the aim of demonstrating phases *within* the network itself in order to understand that there are two regimes of connectivity, i.e. dense and sparse, which demonstrates directly the association of the existence of a second order phase transition with the presence of a power law.

Figure 1 shows the distribution of $n(k)$ (the unnormalized form of $P(k)$) against $k$ for a scale-free network (such as a software network), and Figure 2 shows a plot of Gibbs free energy against inverse temperature for a second-order phase transition in a physical system. We decided to partition the software into two 'phases' (see Figure 1) by comparing network properties with the thermodynamic properties shown in Figure 2. It is already known that power laws most often signal a (second order phase) transition from order to disorder [2]. Power laws are also associated with scale-free networks, so it seems natural to use these to identify the two phases concerned. In Figure 1, $\Gamma_{SF}$, is the 'critical point' analogous to $\Gamma_{GFE}$ where two phases are identical [19] shown in Figure 2. In the case of software networks, no actual matter is involved, but the behaviour can be likened to a 'virtual fluid' (liquid/gas), comprising phase 1 (the 'virtual gas') and phase 2 (the 'virtual liquid') fractions respectively. Phase 1 can be identified with a 'gas' because it contains the nodes that have the lowest density of connections, and the converse is true for phase 2. The liquid-gas phase transition is second-order when crossing the critical point from along the coexistence curve. In addition, for second-order phase transitions, $\partial G/\partial T = -S$ varies *continuously* at constant pressure. The order parameter for the virtual fluid is analogous to the order parameter defined for a real fluid undergoing a second-order phase transition, i.e. the density difference [20].

It can be seen that n($k$), a population, is in effect analogous to energy, and degree $k$ corresponds to 1/T. Consider the Bose-Einstein condensate treatment of a complex network [18] - in our work the network is partitioned into two 'phases' (the analogy is drawn only to explain the mapping between Figure 1 and Figure 2). Phase 2 corresponds to the densely-connected region of Figure 1 where $k$ is large and n($k$) is small i.e. in analogy with the highly populated ground level of the BE condensate [18]. The sparsely populated higher energies correspond to phase 1 where $n(k)$ is large and $k$ is small.

Valverde *et al* [3] were the first group to reduce the structure of software packages to undirected graphs using the concepts of graph theory. We also identify 'classes' (one of the fundamental abstractions used in prevalent modern object-oriented programming languages) with graph nodes, and 'dependencies' [21] (when one class accesses or refers to data or functionality in another class, or has an inheritance relationship) with graph edges. For the purposes of determining whether the software package is associated with a growing network, we consider the graph of software classes (nodes) to be a directed graph. Each node was assigned degrees $k_{in}$, $k_{out}$, and $k_{tot}$, reflecting the number of other nodes which have dependencies on the current node, the number of other nodes on which the current node depends, and the total number of dependencies of the current node, respectively. The rest of the main part of the analysis treats the software as being an undirected graph, i.e. all the analysis that is involved with the partitioning of the software. We used as our test system the 'rt.jar' component from the Sun Java2 Runtime Environment, version 1.4.2_04 [22] where $N$ (total number of nodes (classes)) was 9251. We analysed this component by extracting the node degree statistics and fitting them to a power law. In Figure 1 a linear relation is fitted to the plot of $\log_{10}P_{cum}(k)$ against $\log_{10}k$, where the cumulative distribution $P_{cum}(k)$ is defined as:

$$P_{cum}(k) = \sum_{k'=k}^{\infty} P(k') \qquad (1)$$

This fitting yields a power-law exponent $c$, where $c$ is the gradient of the fitted line. The $\gamma$ exponents for the in, out and total degrees were calculated [16] for the corresponding degree distributions $P(k_{in})$, $P(k_{out})$ and $P(k_{tot})$, the $\gamma$ exponents being given by:

$$1 - c = \gamma \qquad (2)$$

In order to partition the network into two regimes of connectivity in an unbiased fashion we consider the following equality:

$$\sum_{k'=0}^{k_t} k'n(k') = \sum_{k'=k_t}^{k_{max}} k'n(k') \qquad (3)$$

Where $k_{max}$ is the largest value of $k$. Equation (3) is true for a unique degree value $k_t$ which defines the point at which the two 'phases' coexist, (compare $\Gamma_{SF}$ in Fig. 1 and $\Gamma_{GFE}$ in Fig. 2) and the left-hand and right-hand sides define phase 1 and phase 2 respectively. The quantities defined by equation (3) are plotted in Fig. 4. It can be seen from Fig. 4. that the 'phases' meet at a value of $k$ which is within 0.2 % of the location $k_\Gamma$ of the 'knee' of the power law curve as shown in Fig. 1, i.e. where the gradient of the curve is -1. This finding is reasonable, since equation (3) defines a weighted sum, and at $k = k_t$ the increase in the connectivity $k$ of the nodes is approximately balanced by the increase in the number of nodes $n(k)$.

Having partitioned the network into two phases we now consider the most densely connected nodes that comprise phase 2: this is a fraction $Q$ of the total number of nodes. In order to (selectively) test whether removing $Q\%$ of the nodes causes the collapse of a *natural* scale-free network, a test is devised; the network is considered destroyed if $R_t \leq 1$, where $R_t$ and $Q$ are given by the expressions:

$$R(k) = \frac{\sum_{k'=1}^{k} k'}{\sum_{k'=1}^{k} n(k')}, \quad Q = \frac{\sum_{k'=k_{max}}^{k_t} n(k')}{N} \tag{4}$$

The order parameter $R(k)$ varies smoothly which is characteristic of a second order phase transition (see figure 5). The value of $k_t$ obtained from equation (3) was used as an index into the $P_{cum}(k)$ (see equation (1)) data to find $Q$ (see equation (4)), i.e. the percentage of nodes lying at values of $k \geq k_t$. The existence of a network requires that the nodes have sufficient connectivity; a 'radial' cluster of $n_r$ nodes (connected by undirected edges) has a single central node with degree $k = n_r - 1$, and $n_r - 1$ nodes with $k = 1$, each connected only to the central node. The total number of edges for the entire cluster is given by $k_{total} = 2(n_r - 1)$. In order to destroy this cluster, *i.e.* to break it up into an assembly of $n_r - 1$ nodes each with $k = 1$, it is necessary to remove the central node. Once this central node has been removed, it is then trivial to show that the ratio $R = \Sigma k/n_r = 1$ for the remaining fragmented cluster. This is also true for an assembly of radial clusters, since removing the central node always renders $n_r - 1$ separate nodes each with $k = 1$, and if $R = 1$ we can be sure that all of the clusters have been fragmented. Using this additional fragmentation criterion, we can investigate the increase in efficiency gained by moving from $Q$ to $Q_{lim}$ (the smallest possible number of nodes required to hold a network together) by choosing $R = 1$ as a boundary condition. The new measure $Q_{lim}$ applies for $R = 1$, where $k_{lim}$ is defined by $R(k_{lim}) = 1$, from equation (4),

$$Q_{lim} = \frac{\sum_{k'=k_{max}}^{k_{lim}} n(k')}{N} \tag{5}$$

To summarize the thermodynamics analogy; the pure 'liquid' state corresponds to a value of the order parameter (analogous to the density) $R(k) \geq 1$, for a second-order gas/liquid phase transition. For very small $R(k)$ (i.e. as close as possible to zero) the connectivity of the network is 'gas-like' and for the region close to $k_t$, the network is a mixture of 'gas' and 'liquid'- like 'phases'.

For the 'rt.jar' component used in this study, $Q = 9.4\%$, and $Q_{lim} = 1.1\%$. From our analysis, we postulate that all of the nodes associated with phase 2 should be given priority in testing (i.e. those lying within the fraction $Q$), with the subset $Q_{lim}$ of $Q$ representing the absolute minimum subset of nodes that must be thoroughly tested. The magnitude of $Q$ correlates well with the findings in the software industry i.e. that 10 % of the nodes are responsible for 90% of the downtime and $Q_{lim}$ with the 2% of the "Dead in the Water" defects. This shows that errors on the highly connected nodes are troublesome, since when repaired their highly-connected character means that the surrounding nodes to which they are connected must also be updated. This selective-testing approach will be of use to software engineers, who will find it useful to know which and how many of the classes need to be prioritised for tested, without relying on chance, i.e. using random testing or wasting time by exhaustive testing. These classes need to be tested for errors early in a project, in order to minimize later retesting efforts as the software grows. The software investigated in this study can be seen to be growing, since $\gamma_{Out} > \gamma_{In}$ (see Figure 3 and associated caption).

These results of this study will also be widely applicable to other scale-free complex networks, since all natural networks considered to date in the literature have possessed homeostasis against random failure [2]. Silica aerogels have been found to have the same fractal dimensions right down to near-molecular scales [23], as revealed by small angle neutron scattering i.e. they are practically scale-free. It was found [23] that the silica aerogel scattering data obeyed power law fits of correlation length versus density. It is now possible using network theory to simply evaluate that the value of $\gamma$ (using equation (2) applied to the results stated in [23] of the power law fits of the scattering) for the sample of silica aerogel was $2.67 \pm 0.05$. In future it will be possible to test the hypothesis that the quantity $Q_{lim}$, corresponds to the smallest possible number of nodes (structural subgroups) that need to be removed in order to disassemble natural scale-free networks in a similar manner to the work of Albert *et al* [17]. In doing so it will be possible to obtain the theoretical limit on the minimum density of silica aerogels.

This work can be used to develop a new method to theoretically model other scale-free materials and networks that do possess homeostasis against random failure. Other possible applications include finance, where it may be possible to find the most efficient way to spend a marketing budget using *only* those nodes required to percolate the entire network; this would simply mean identifying the $Q$ and $Q_{lim}$ nodes. The fragmentation of natural networks after the loss of the few highly connected nodes [24] which maintain the networks connectivity can be considered in the case of silica aerogel networks to be associated with the onset of a catastrophic structural failure. In summary, we have demonstrated that it is possible to partition scale-free networks into phases by treating the network as a fluid close to a second order phase transition. It is hoped that this work provides some insights into the leading questions for network research such as 'providing insights on the structure and classification of complex networks in precise mathematical terms' [25].

**Acknowledgements**

This work was supported by the KK foundation (grant no 2001/204).

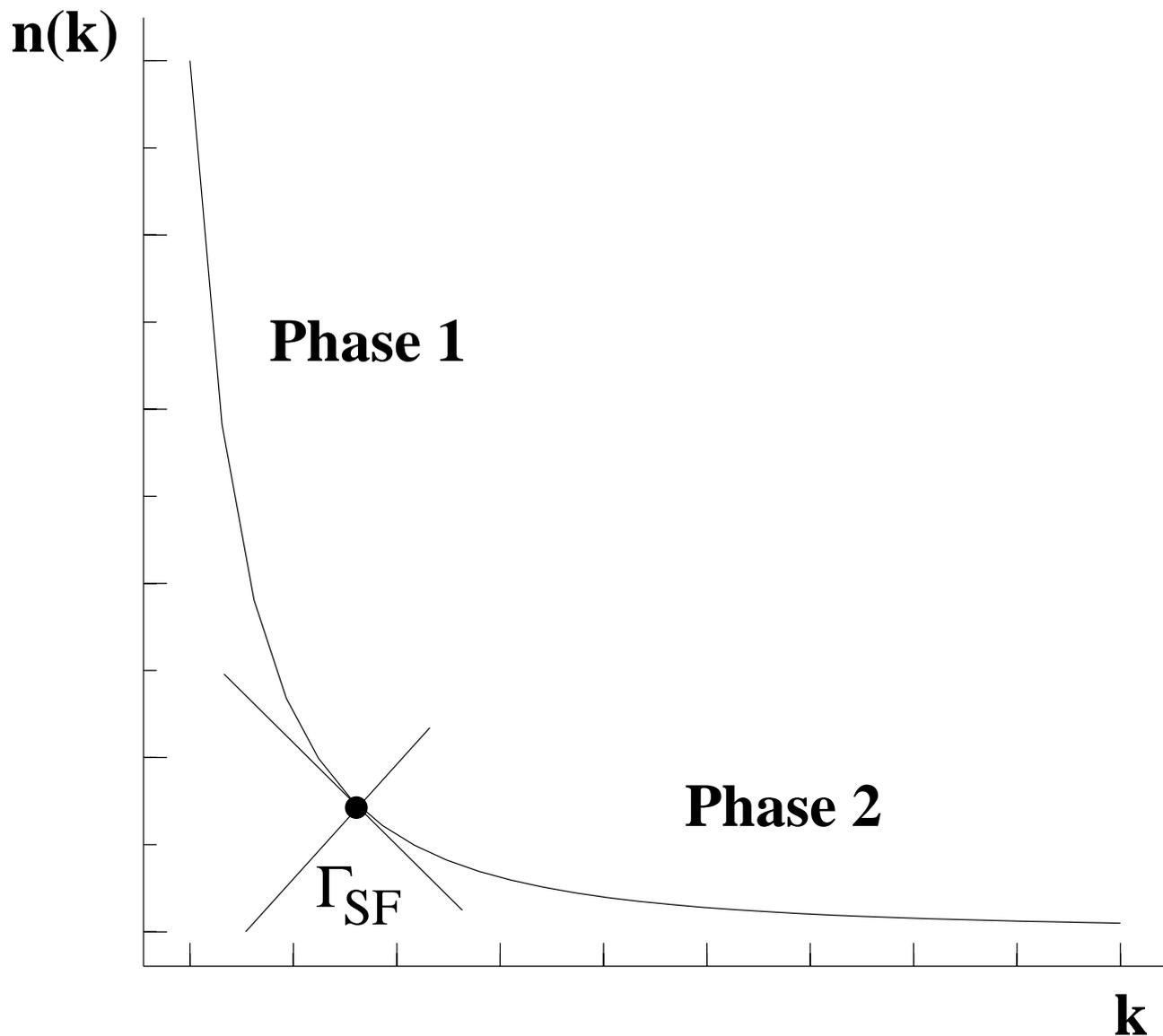

Fig. 1 - A diagram of the dependency of the number of nodes *n(k)* with connectivity *k* for a general scale-free network. The position $\Gamma_{SF}$ corresponds to the position where the gradient equals $-1$.

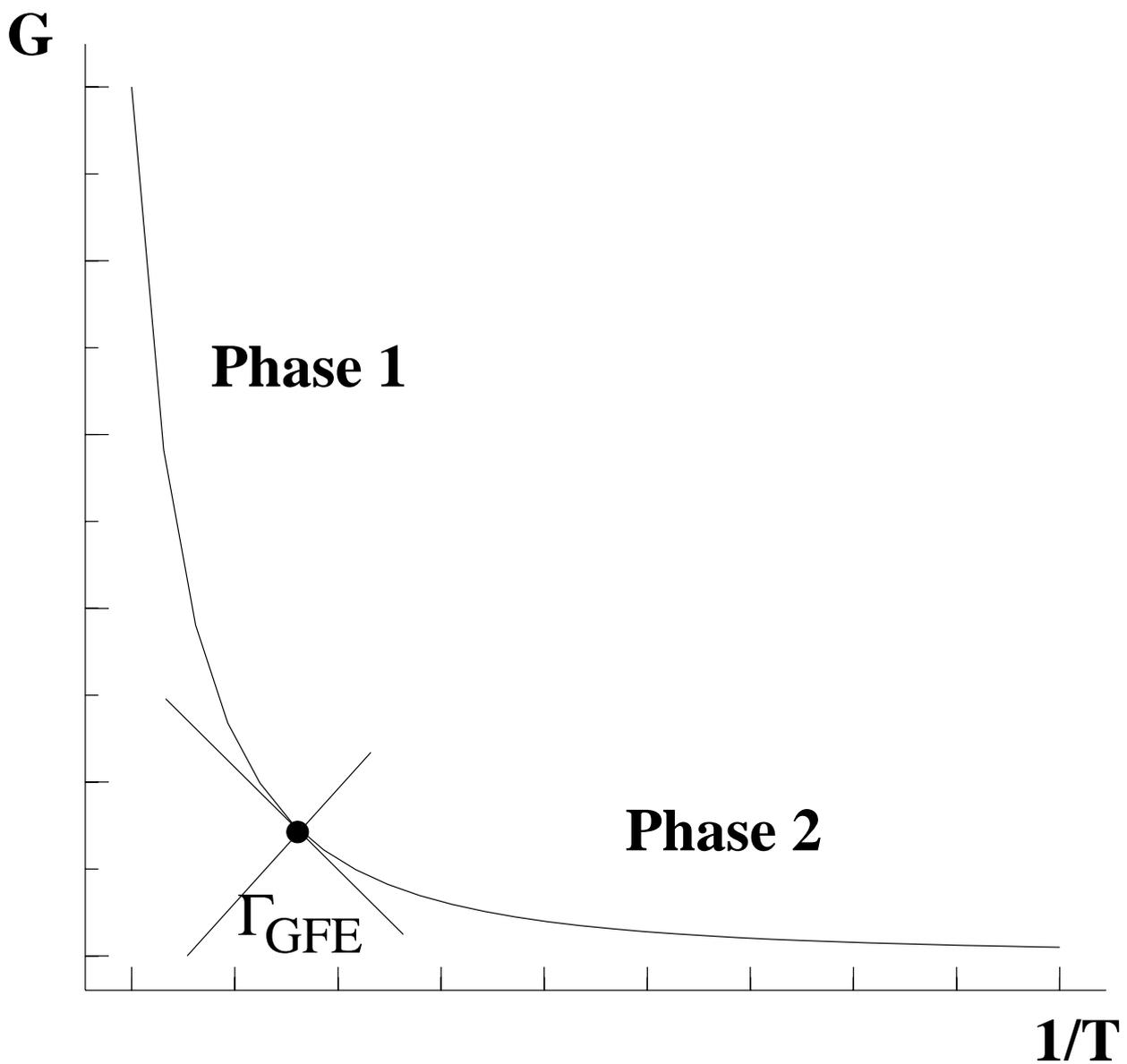

Fig. 2 – A diagram of the dependency of the Gibbs free energy on inverse temperature in a physical system. The position $\Gamma_{GFE}$ corresponds to the co-existence of two phases in the Gibbs free energy close to a second order phase transition.

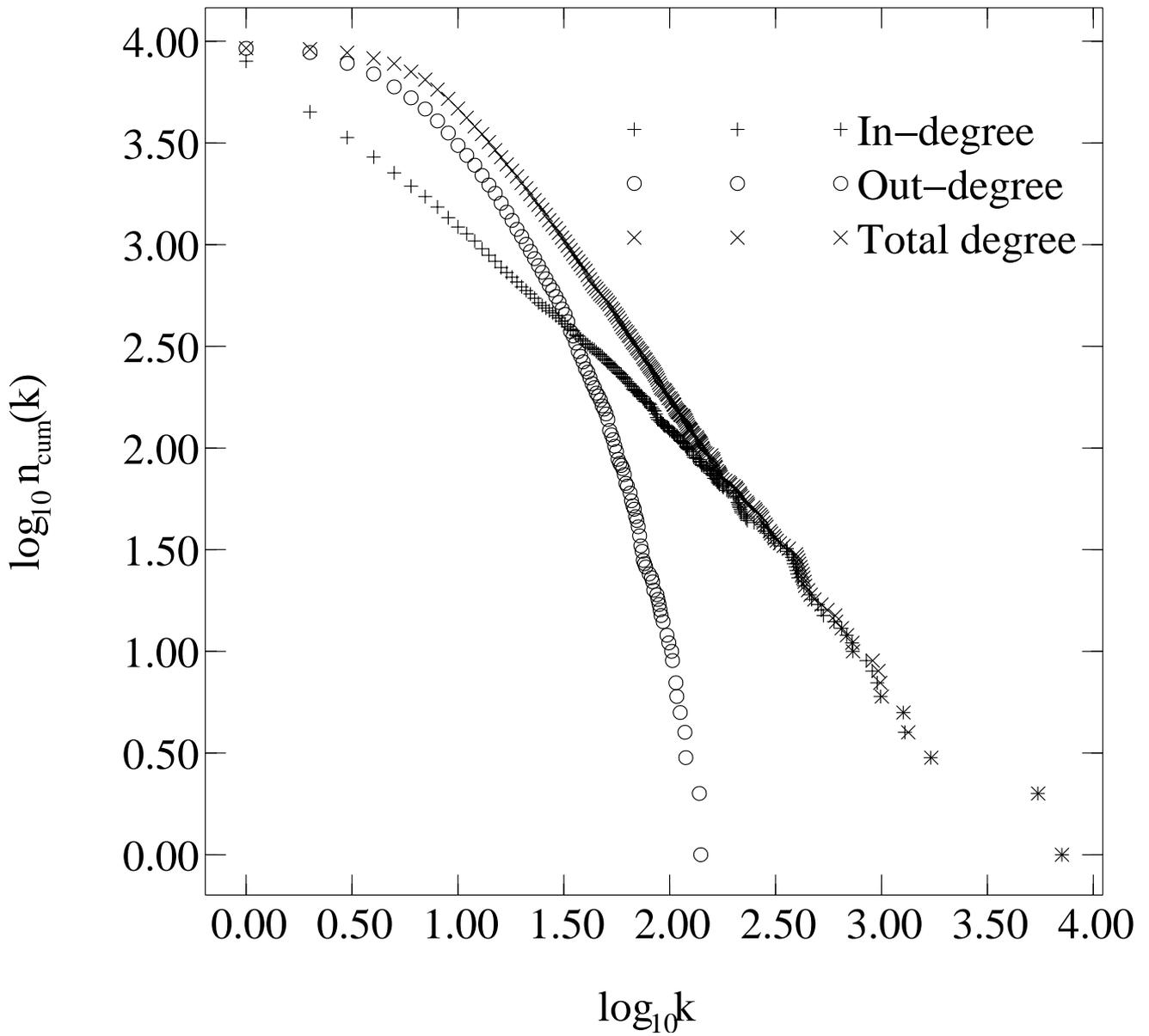

Fig. 3 -The power law fits give $\gamma_{In}$ = 2.15, $\gamma_{Out}$ = 3.69, $\gamma_{Total}$ = 2.46 for the -in, -out and total degrees, the correlations of these fits were 0.99, 0.93, 0.99 respectively.

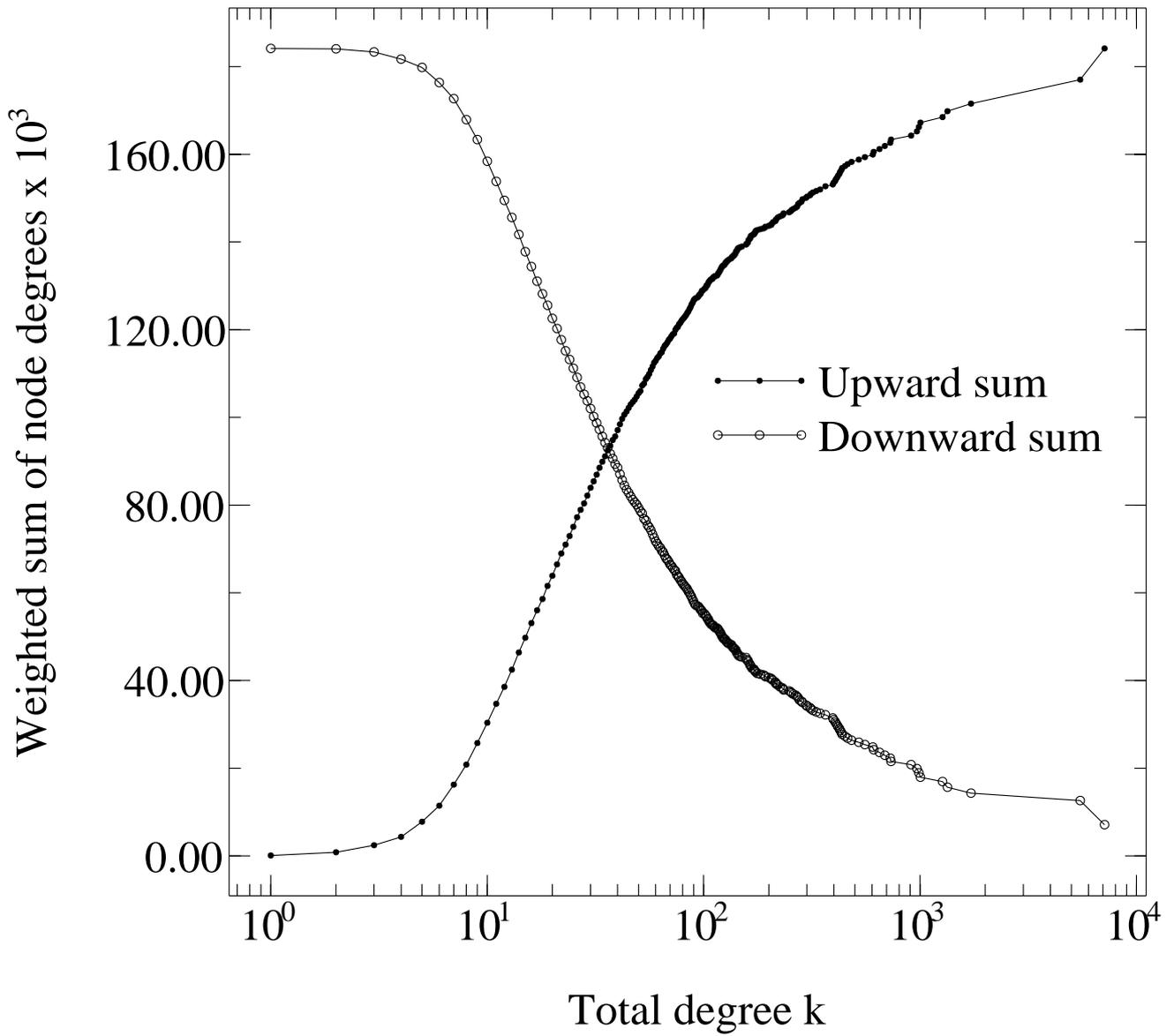

Fig. 4 - Graphs of the weighted sum of nodes degrees $n(k)$ vs. the total degree $k$. In graph legend the labels upward and downward sums refer to the left and right hand sides of equation (3) respectively.

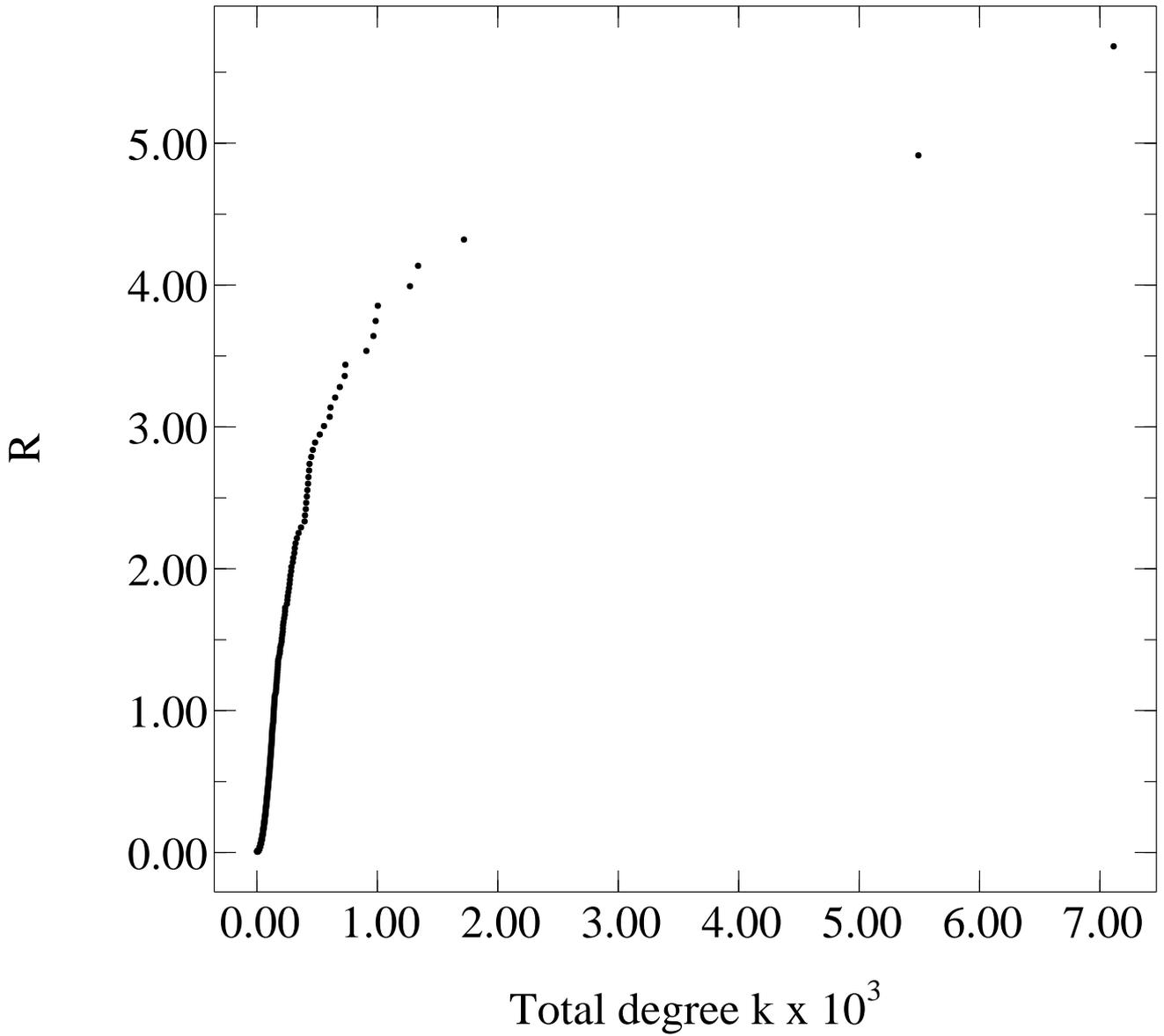

Fig. 5 -A plot of the order parameter $R(k)$, see equation (4) for the definition.